\documentstyle{l-aa}

\begin{document}

  \thesaurus{
													(02.07.1;
              02.08.1;
              02.09.1;
              11.06.1;
              12.04.1)}

\title{On the onset of the Jeans instability in a two-component fluid}  

\author{J. P. M. de Carvalho\inst{} 
\and P. G. Macedo\inst{}}

\offprints{J. P. M. de Carvalho}

\institute{Grupo de Cosmologia do C.A.U.P.
\thanks{A member of the European Cosmology Network}\\
R. do Campo Alegre, 823\\
4100  Porto - Portugal}

\date{}

\maketitle

\begin{abstract}

Conditions for the establishment of small density perturbations in a
self-gravitating two component fluid mixture are studied using a dynamical
system approach.

It is shown that besides the existence of exponentially growing and decaying
modes, which are present for values of the perturbation wave-number $k$ smaller
than a critical value $k_{_M}$, two other, pure oscillatory, modes exist at all
scales.  For $k < k_{_M}$, the growing mode always affects both components of
the fluid and not only one of them.

Due to the existence of a resonance between the baryonic and the dark
perturbations, it is shown that the onset of structure formation in the post
recombination epoch is substantially enhanced in a narrow scale band around
another critical value $k_{c}$. For dark matter particles having a mass $\sim
30$ eV, the corresponding critical mass scale for the establishment of density
perturbations at the time of recombination is of the same order of magnitude as
the galactic one. 


\end{abstract}

\section{Introduction}

	The study of the mechanisms responsible for the formation of structures in the universe was
started by the pioneer work of Jeans (1902, 1928). 
	Studying the conditions under which a cloud of gas (in a static background) becomes
gravitationally unstable under its own gravity Jeans concluded that perturbations with masses
smaller than a "critical" value $M_{_J}$ (the Jeans mass) do not grow and behave like
acoustic waves whereas perturbations with a mass $M$ bigger than $M_{_J}$ grow under
the effect of their self-gravity, leading to gravitationally bound structures.

 Since then, the mechanism proposed by Jeans has been extensively applied as a criterion of
stability in several models of galaxy formation.
	However, by the recombination epoch (which is believed to be the one after which baryonic
perturbations could begin to grow), the critical scale predicted by the classical Jeans
theory is not at all related to the galactic scale. 
	Instead, it is well known that just before recombination the critical Jeans mass is $\sim
10^{16-17}$M$_{\odot}$, (i.e. the mass of rich clusters or even superclusters of galaxies),
and immediately after recombination the Jeans mass drops abruptly, more than 10 orders of
magnitude, to a value $\sim 10^{5-6}$M$_{\odot}$, which is related to the mass of globular
clusters (Weinberg 1972; Zel'dovich \& Novikov 1983). 
 
	The origin of this discrepancy between the Jeans mass, before and after recombination, and
the mass of a typical galaxy, a few $10^{11}$M$_{\odot}$, has not been, up to now, well
understood. 

	One should emphasize, however, that these values for the Jeans mass just before and just
after recombination epoch are obtained assuming a baryon-dominated universe.

 On the other hand, the observation of flat galactic rotation curves at large galactic radii
as well as high galaxy velocity dispersions in clusters has led to the missing mass problem and
to the dark matter conjecture to solve it.
 This fact has, in particular, led to the hypothesis that dark matter halos exist around
galaxies.

	The existence of dark matter therefore also implies that when studying the dynamics and the
formation of structures in the universe one cannot use the simple Jeans stability criterion
for a one component gas, but one has to study how does the gravitational instability arise in
a mixture of at least two components.
	This formulation of the problem leads to a different Jeans mass, as well as to a better
understanding of the origin of galactic mass spectrum.
 This will be the main goal of this paper.

	We must note that, after we finished this work, it was pointed out to us by Dr. Varun Sahni
that this problem had already been studied, using a different formalism, by some russian
authors like Grishchuk \& Zel'dovich (1981) or Polyachenko \& Fridman (1981). 
	Although having reached some of our conclusions, such studies lack most of the features
present in this work. In particular they do not refer to the existence of the resonance we
shall point out in this paper.

	Since it is generally accepted that dark matter is constituted by WIMPs, (Weakly
Interacting Massive Particles) in this study we shall consider the case for which the two
components of the cosmic fluid interact only gravitationally.

	The two fluid components that we shall consider will be a baryonic one and a hot dark matter
(HDM) one, hereafter component $B$ and $D$ respectively. 
 The latter being assumed to be made of neutrino-like particles. 

 Although the numerical results obtained in this paper refer to the particular case of HDM,
one should point out that the formalism here developed is quite general and applies to any
type of a two-component mixture.

	In Sect. \ref{sec.equ} we shall establish the linear coupled equations which describe the
dynamical system for the density perturbations in the two-component fluid we wish to study.
	We shall then develop, in Sect. \ref{sec.qual}, a qualitative analysis of the dynamical
system.

	In Sect. \ref{sec.sol} we shall study the behaviour, in phase-space, of the trajectories
representing the general solution of the dynamical system. 
 This analytical study will be complemented with some numerical plots in order to clarify some
of the qualitative results obtained.
 
	For numerical purposes we shall consider that the onset of gravitational instability occurs
in a flat $(\Omega = 1)$ universe at recombination epoch, after the decoupling between
radiation and baryonic matter, when the radiation temperature is $\sim 3000$ K (Kolb \&
Turner 1990). 
	We shall assume that the density parameter $\Omega_B$ has a value in agreement with the
limits imposed by standard nucleosynthesis (Walker et al. 1991). 
 Assuming $0.5 \leq$ h $\leq 1$, where h is the Hubble constant in units of $100$ Km/s/Mpc,
i.e. h = H$_0/$($100$ Km/s/Mpc), such a value is $\sim 0.05$.

	The neutrino-like particles which constitute the dark matter component are assumed to have a
mass compatible with the assumption of being already non-relativistic by the recombination
epoch. 
	It will be shown later in this paper that this value for the mass of the dark matter
particles seems to provide a good fit for the galactic mass spectrum. 

	In Sect. \ref{sec.res} we shall describe the occurence of a resonance which, in our
opinion, is responsible for the formation of structure at the typical galactic scales.

	The relation between the existence of the resonance, the galactic scale and the mass of dark
matter particles is discussed in Sect. \ref{sec.gxf}.

	Finally in Sects. \ref{sec.con} and \ref{sec.fut} we shall point out the main results
presented in this paper and outline some open questions as well as the research paths we
intend to follow in the near future.

\section{Dynamical equations for a two component fluid} \label{sec.equ}

	The dynamics of a non-relativistic self-gravitating fluid can be described by the usual set
of hydrodynamical equations: continuity equation, Euler equation and Poisson's equation.

	In this paper we shall neglect the expansion of the universe, since we shall be interested
only in the qualitative behaviour, at the onset of the process of growth of density
perturbations in the post recombination epoch, and not in the exact dynamics of the process.
 
 In this case, for a fluid made of two components, $D$ and $B$, the above set of equations
becomes:
 
\begin{equation}
\frac{\partial \rho_{_D}}{\partial t}+\nabla \mbox{\boldmath
$\cdot$} \, (\rho_{_D} \mbox{\boldmath $v$}_{_D})=0
\label{eq.A} 
\end{equation}

\begin{equation}
\frac{\partial \rho_{_B}}{\partial t}+\nabla \mbox{\boldmath
$\cdot$} \, (\rho_{_B} \mbox{\boldmath $v$}_{_B})=0
\label{eq.B} 
\end{equation}

\begin{equation}
\frac{\partial \mbox{\boldmath $v$}_{_D}}{\partial t}+(\mbox{\boldmath $v$}_{_D} \mbox{\boldmath
$\cdot$} \, \nabla) \, \mbox{\boldmath $v$}_{_D} =
-\frac{1}{\rho_{_D}}\nabla p_{_D}-\nabla \Phi 
\label{eq.C}
\end{equation}

\begin{equation}
\frac{\partial \mbox{\boldmath $v$}_{_B}}{\partial t}+(\mbox{\boldmath $v$}_{_B} \mbox{\boldmath
$\cdot$} \, \nabla) \, \mbox{\boldmath $v$}_{_B} =
-\frac{1}{\rho_{_B}}\nabla p_{_B}-\nabla \Phi 
\label{eq.D}
\end{equation}

\begin{equation}
\nabla^{2} \Phi = 4\pi G \, (\rho_{_D} + \rho_{_B}) 
\label{eq.E}
\end{equation}
where $\Phi$ is the total self-gravitational potential, $\rho_j$, $\mbox{\boldmath $v$}_j$ and
$p_j$ are respectively the density, velocity and pressure fields for fluid elements of
component $j$, (the index $j$ taking the values $D$ or $B$) .

	Using a perturbative analysis, to first order in the perturbation, one can write the
dynamical variables as follows:

\begin{equation}
 \begin{array}{l}
 \rho_{_j} = \rho_{_j{_{_0}}} + \delta \rho_{_j} \\
 \mbox{\boldmath $v$}_{_j} = \mbox{\boldmath $v$}_{_j{_{_0}}} + \delta \mbox{\boldmath
$v$}_{_j} \\
 p_{_j} = p_{_j{_{_0}}} + \delta p_{_j} \\
 \Phi_{_T} = \Phi_{_T{_{_0}}} + \delta \Phi_{_T}
 \end{array} 
\end{equation}
where the index {\small "$0$"} indicates the unperturbed value, and $\delta$ the first order
perturbation.
 
 It's common to consider the zero order solution to be an infinite fluid at rest,
($\mbox{\boldmath $v$}_{_0} = 0$), for which the density and pressure are the same everywhere
(unperturbed fluid).
	We must point out, as it is well known, that this static solution for the unperturbed state
of the above equations is not mathematically correct, except for $\rho_{_0} = 0$. 
	However, since this problem is removed when considering the more realistic case of a fluid
embedded in the expanding universe, (which we shall discuss in a forthcoming paper), one can
start perturbing this static configuration in order to obtain some qualitative information on
the fluid behaviour.

 We shall assume that the solutions of the above coupled system of Eqs. (\ref{eq.A}) -
(\ref{eq.E}) are square integrable functions of the spatial variables and therefore can be
Fourier analysed with respect to these variables. 
	
	To first order in the perturbation, after the Fourier analysis in the spatial variables and
some manipulation, the set of Eqs. (\ref{eq.A}) - (\ref{eq.E}) leads to the following
equations:
 
\begin{equation}
\left\{
\begin{array}{lll}
\ddot{\Delta}_{_D} + \left( c_{_D}^{2} \, k^{2} - W \! {_{_D}} \right) \Delta_{_D} - W \!
{_{_B}} \, \Delta_{_B} &=& 0   
\label{eq.I} \\
\\   
\ddot{\Delta}_{_B} + \left( c_{_B}^{2} \, k^{2} - W \! {_{_B}} \right) \Delta_{_B} - W \!
{_{_D}} \, \Delta_{_D} & = & 0
\end{array}
\right.   
\label{eq.J}   
\end{equation} 
 where the double dot stands for second time derivative, $k$ is the wave number of the
perturbation. For each of the components $j$, the parameter $W \! _{j}$ is given by $W \! _{j}
= 4 \pi G \rho_{_j{_{_0}}}$ and $\Delta_{j}$ is the density contrast $\delta \rho_{_j}  /
\rho_{_j{_{_0}}}$ and $c_j$ the adiabatic sound speed given by $c_{_j} = \left( \partial
p_{_j} / \partial \rho_{_j} \right)^{1/2}$. 

	The value of the parameter $W \! _{j}$, at the cosmic time of recombination, $t_{rec}$, can
be related to the density parameter, $\Omega_j$, of component $j$, by

\begin{equation}
W \! _{j} = \frac{\,2\,}{3} \, \frac{ \Omega_j } { t_{rec}^2 } 
%\label{eq.14}
\end{equation}

	The adiabatic sound speed in component $j$ at recombination, $c_{j_{rec}}$, is the sound
speed of a monoatomic ideal gas, which is given by:

\begin{equation}
c_{j_{rec}} = \left( \frac{\,5\,} {3} \, \frac{ K_{_{Bol}} T_{j_{rec}} } { m_{j} } \right)^{1/2}
 \label{eq.13}
\end{equation}
where $K_{_{Bol}}$ is the Boltzman constant and $T_{j_{rec}}$ and $m_{j}$ are respectively the
temperature at recombination and mass of the particles of component $j$.

	One should emphasize, that from the decoupling between the neutrino-like particles and
radiation until the epoch at which these particles become non-relativistic, (which occurs
for a red-shift $z_{_{NR}} \sim 6 \times 10^4$, Doroshkevich et al. 1988), their temperature
T$_{D}$ is given by T$_{D} = (4/11)^{1/3}\,$T$_{\gamma}$, where T$_{\gamma}$ is the radiation
temperature. 
 Since then, it decays with $a^{-2}$, where $a$ is the scale factor of the universe
(Gao \& Ruffini 1981; Padmanabhan 1993).
	One can therefore estimate that, at the recombination epoch, such a dark component $D$ will
have a temperature T$_{D{_{rec}}} \sim 40$ K 

 The above system (\ref{eq.J}) of two linear differential equations of second order can be
transformed into the following autonomous dynamical system of four first order differential
equations which one can write in a matrix form as:

\begin{equation}
\dot{\mbox{\boldmath $x$}} = \mbox{\boldmath $A \, x$}
\label{eq.1}
\end{equation}  
where $\mbox{\boldmath $x$} \equiv (x_1 , x_2 , x_3 , x_4)^{\rm T} \equiv (\dot{\Delta}_{_D} ,
\Delta_{_D}, \dot{\Delta}_{_B} , \Delta_{_B})^{\rm T}$, the superscript ${\rm T}$ standing for
transpose. $\mbox{\boldmath $A$}$ is a matrix whose
components are only functions of the parameter $k$ and has the form: 
 
\begin{equation}
\mbox{\boldmath $A$} \equiv \mbox{\boldmath $A$}(k) = \left(
\begin{array}{cccc}
0 & W \! {_{_D}} - c_{_D}^{2} \, k^{2} & 0 & W \! {_{_B}} \\
1 & 0 & 0 & 0 \\
0 & W \! {_{_D}} & 0 & W \! {_{_B}} - c_{_B}^{2} \, k^{2} \\
0 & 0 & 1 & 0
\end{array} 
\right)
\label{eq.2}
\end{equation}  

	We shall be particularly interested in the solutions which represent the density perturbations
in components $D$ and $B$. 
	These solutions are given by the second and fourth components, $x_2(t)$ and $x_4(t)$, of the
general solution $\mbox{\boldmath $x$}(t)$, of the dynamical system (\ref{eq.1}).

	Using methods of qualitative theory of differential equations one can analyse the dynamical
system and obtain some qualitative information on the behaviour of the perturbations. This
will help understanding the stability, against density perturbations, of the
two-component system. Even in a case like this one where an explicit solution exists, this
is worth doing.

	Alternatively the autonomous dynamical system can be numerically studied provided we specify the
initial condition $\mbox{\boldmath $x$}_0 \equiv \mbox{\boldmath $x$}(t_0)$, where $t_0$ is the
cosmic time at which the equilibrium state of the fluid is perturbed.
	
	We shall analyse the system qualitatively and, whenever it proves convenient, to clarify any
result, we shall illustrate this study with some numerical plots.

\section{Qualitative study of the dynamical system} \label{sec.qual}

 The above matrix $\mbox{\boldmath $A$}(k)$ has $4$ different eigenvalues for all values of
$k$ except for $k = 0$ and for a critical value $k = k_{_M}$, which shall be defined below.
 Therefore for $k > 0$ and $k \neq k_{_M}$ there are $4$ linearly independent eigenvectors. 
	In this case the general solution $\mbox{\boldmath $x$}(t)$ of the system (\ref{eq.1}) is
well known (Arnold 1973).

 If all the eigenvalues are real ones, that solution can be written in the form of a linear
expansion of particular solutions ${\rm e}^{\lambda_j t} \, \mbox{\boldmath $\xi$}_j$, i.e.:

\begin{equation}
\mbox{\boldmath $x$}(t) = \sum_{j=1}^{4} \alpha_{j} \, {\rm e}^{\lambda_j t} \,
\mbox{\boldmath $\xi$}_j
 \label{eq.1-a}
\end{equation}
where $\alpha_{j}$ are real-valued functions of the the parameter $k$, which are determined by
the initial conditions, and $\lambda_j$ and $\mbox{\boldmath $\xi$}_j$ are respectively the
eigenvalues of the matrix $\mbox{\boldmath $A$}(k)$ and its associated eigenvectors. 

	If some of the eigenvalues are complex conjugates, one can easily see that the real and
imaginary parts of the complex function ${\rm e}^{\lambda_j t} \, \mbox{\boldmath $\xi$}_j$
are two real-valued particular solutions of the dynamical system associated with the complex
conjugate pair of eigenvalues $\lambda_j$ and $\overline{\lambda}_j$.  
  
 In order to obtain some qualitative insight about the behaviour of the perturbations, one
must find the critical or equilibrium points of the system (\ref{eq.1}) and calculate the
eigenvalues of matrix $\mbox{\boldmath $A$}(k)$ as well as their associated eigenvectors.

\subsection{Critical points, eigenvalues and eigenvectors}

	The critical points of the dynamical system (\ref{eq.1}) are obtained solving the equation 

\begin{equation}
\dot{\mbox{\boldmath $x$}} = 0. 
\label{eq.2-a}
\end{equation}

	The only critical point which does not depend on the parameter $k$ is the origin,
$\mbox{\boldmath $x$} \equiv (0,0,0,0)^{\rm T}$. 

	The parameter-dependent critical points are given by the solutions of the equation
$\det(\mbox{\boldmath $A$}) = 0$. 
 In the case when the parameter $k$ has a value given by:

\begin{equation}
k_{_M}^{2} = k_{_B}^{2} + k_{_D}^{2} = \frac{W \! {_{_B}}}{c_{_B}^{2}} + \frac{W \!
{_{_D}}}{c_{_D}^{2}} \; , 
 \label{eq.3}
\end{equation}
any point $\mbox{\boldmath $x$} \equiv (0,x_{_2},0,x_{_4})^{\rm T}$ of the $(x_2 , x_4)$ plane,
is a critical point of the above system.
 
 In Eq. (\ref{eq.3}), $k_{_B}$ and $k_{_D}$ are respectively the Jeans wave numbers of
components $B$ and $D$ when taken separately, and we shall call $k_{_M}$ the Jeans wave
number for the mixture of the two components.

	The eigenvalues of the matrix $\mbox{\boldmath $A$}(k)$ are the roots of the characteristic
polynomial, obtained solving the equation

\begin{equation}
\det(\mbox{\boldmath $A$} - \lambda \,
\mbox{\boldmath $I$}) = 0
\label{eq.3-a}
\end{equation}
where $\mbox{\boldmath $I$}$ is the identity matrix.

	The 4 eigenvalues of $\mbox{\boldmath $A$}(k)$, which we shall represent by $\lambda_{\,1}$,
$\lambda_{\,2}$, $\lambda_{\,3}$ and $\lambda_{\,4}$, are therefore given by:

\begin{equation}
\left\{
\begin{array}{lllll}
\lambda_{1} & = & - \, \lambda_{2} & = & \frac{1}{\sqrt{2}} \sqrt{f + \sqrt{f^{2} + 4k^{2}g}} \\
 & & \\
\lambda_{3} & = & - \, \lambda_{4} & = & \frac{1}{\sqrt{2}} \sqrt{f - \sqrt{f^{2} + 4k^{2}g}} \\ 
\end{array}
\right.
\label{eq.4}
\end{equation}
where $f$ and $g$ are the following functions of $k$:

\begin{equation}
f(k) = W \! {_{_B}} + W \! {_{_D}} - k^{2} \left( c_{_B}^{2} + c_{_D}^{2} \right)   
\label{eq.6}
\end{equation}

\begin{equation}
g(k) = W \! {_{_B}}c_{_D}^{2} + W \! {_{_D}}c_{_B}^{2} - k^{2} c_{_B}^{2} c_{_D}^{2} 
\label{eq.7}
\end{equation}

 The eigenvectors $\mbox{\boldmath $\xi$}_j$ of the matrix $\mbox{\boldmath $A$}(k)$, associated to
the eigenvalues $\lambda_j$, must satisfy the condition 

\begin{equation}
\mbox{\boldmath $A$} \, \mbox{\boldmath $\xi$}_j = \lambda_j \, \mbox{\boldmath $\xi$}_j,
\label{eq.7-a}
\end{equation}
and can be written in the following form: 

\begin{equation}
\left\{
\begin{array}{lll}
 \mbox{\boldmath $\xi$}_{_1} & \equiv & (\beta_{1} \, \lambda_{1}, \; \beta_{1}, \; \lambda_{1}, \;
1)^{\rm T} \\ 
 \mbox{\boldmath $\xi$}_{_2} & \equiv & (\beta_{2} \, \lambda_{2}, \; \beta_{2}, \; \lambda_{2}, \;
1)^{\rm T} \\ 
 \mbox{\boldmath $\xi$}_{_3} & \equiv & (\beta_{3} \, \lambda_{3}, \; \beta_{3}, \; \lambda_{3}, \;
1)^{\rm T} \\ 
 \mbox{\boldmath $\xi$}_{_4} & \equiv & (\beta_{4} \, \lambda_{4}, \; \beta_{4}, \; \lambda_{4}, \;
1)^{\rm T} \\ 
 \end{array}
 \right.
 \label{eq.8}
\end{equation}
where $\beta_{j}$ are also functions of $k$ and are given by:
 
\begin{equation}
\left\{
\begin{array}{lllll}
\beta_{1} & = & \beta_{2} & = & \frac{1}{2 W \! {_{_D}}} \left( h \, + \, \sqrt{h^{2} + 4W
\! {_{_B}}W \! {_{_D}}} \; \right) \\
 &  &  \\
\beta_{3} & = & \beta_{4} & = & \frac{1}{2 W \! {_{_D}}} \left( h \, - \, \sqrt{h^{2} + 4W
\! {_{_B}}W \! {_{_D}}} \; \right) \\ 
\end{array}
\right.
\label{eq.9}
\end{equation}
and $h = h(k)$ is given by:

\begin{equation}
h(k) = W \! {_{_D}} - W \! {_{_B}} + k^{2} \left( c_{_B}^{2} - c_{_D}^{2} \right)
\label{eq.10}
\end{equation}

\subsection{Qualitative results}

	Qualitative information about the behaviour of the dynamical system can be obtained studying the
signs of the real and imaginary parts of the eigenvalues $\lambda_{j}$.  

	The only zero of the eigenvalues $\lambda_{\,1}$ and $\lambda_{\,2}$ is obtained when $k$ is the
Jeans wave-number of the mixture, ($k = k_{_M}$). 

	Looking at Eqs. (\ref{eq.4}) - (\ref{eq.7}) one can see that for $k < k_{_M}$ the function
$g(k)$ is positive and therefore $\lambda_{\,1}$ and $\lambda_{\,2}$ are real nonzero
eigenvalues, ($\lambda_{\,1} > 0$, $\lambda_{\,2} < 0$). 
 For $k > k_{_M}$, $\lambda_{\,1}$ and $\lambda_{\,2}$ are two pure imaginary conjugate
eigenvalues.

 The eigenvalues $\lambda_{\,3}$ and $\lambda_{\,4}$ do not have zeros, except for the
asymptotic value $k = 0$. 
 Since $f - \sqrt{f^{2} + 4k^{2}g}$ is always negative, for all other values of $k$ 
$\lambda_{\,3}$ and $\lambda_{\,4}$ are pure imaginary conjugate eigenvalues.

	Physically, this means that one has two paired wave modes (which correspond to acoustic
oscillations), for all values of the wave number $k$ and two other modes which are acoustic
for $k > k_{_M}$ and become one growing and one decaying mode for $k < k_{_M}$. 

 These results are summarized in Table \ref{tb.1}.

\begin{table}[t]
 \caption[]{The signs of real and imaginary parts, ($\Re$ and $\Im$ respectively), of the four
eigenvalues of matrix $A(k)$ as a function of parameter $k$. Note that $\Im(\lambda_{3}) = -
\Im(\lambda_{4}) \neq 0$ for all $k \neq 0$. (See text for discussion.)}  
 \label{tb.1}
\begin{flushleft}
%\begin{center}
\begin{tabular}{llcccc} 
\noalign{\smallskip}
\hline
\noalign{\smallskip}
 &   & $k = 0$ & $0 < k < k_{_M}$ & $k = k_{_M}$ & $k > k_{_M}$ \\ 
\noalign{\smallskip}
\hline\noalign{\smallskip}
 $\lambda_{1}$ & $\Re$ & $+$ & $+$ & $0$ & $0$ \\  
               & $\Im$ & $0$ & $0$ & $0$ & $+$ \\ 
 \\
 $\lambda_{2}$ & $\Re$ & $-$ & $-$ & $0$ & $0$ \\ 
               & $\Im$ & $0$ & $0$ & $0$ & $-$ \\ 
 \\  
 $\lambda_{3}$ & $\Re$ & $0$ & $0$ & $0$ & $0$ \\ 
               & $\Im$ & $0$ & $+$ & $+$ & $+$ \\ 
 \\
 $\lambda_{4}$ & $\Re$ & $0$ & $0$ & $0$ & $0$ \\ 
               & $\Im$ & $0$ & $-$ & $-$ & $-$ \\
\noalign{\smallskip}
\hline
\end{tabular}
%\end{center}
\end{flushleft}
\end{table}

	One can therefore distinguish two, very distinct, situations concerning the four-dimensional
phase-space according to the values of the parameter $k$, i.e.:

\begin{enumerate}

\item For $k > k_{_M}$ the four eigenvalues $\lambda_1$, $\lambda_2$, $\lambda_3$ and
$\lambda_4$, are pure imaginary and therefore the phase-space is a {\em center space}, which
is spawned by the eigenvectors associated to those eigenvalues. 

 Solutions in a center subspace oscillate at constant amplitude, and therefore density
perturbations with wave number $k > k_{_M}$ can't grow.
	This case corresponds to the existence of two pairs of acoustic wave modes propagating in the
fluid with different phase velocities. 
	In the limit of small wavelengths, it is easy to see that these modes have the same
dispersion relations and compositions as two completely uncoupled acoustic waves propagating
in pure $D$ and $B$ fluids. 
	This is due to the fact that the role  played by their self gravity, which is the only
mechanism coupling the two components, is not important at small scales.      

\item For $k < k_{_M}$, (see Table \ref{tb.1}), the eigenvalue $\lambda_1$ is real and positive,
$\lambda_2$ is real and negative and $\lambda_3$ and $\lambda_4$ are pure imaginary
(conjugates) eigenvalues. 
 In this case the phase-space is, therefore, the direct product of three distinct subspaces: 

\begin{enumerate}

\item An {\em unstable subspace} of dimension one spawned by the eigenvector associated to the
      eigenvalue with positive real part, ($\lambda_1$).

			   Solutions in an unstable subspace always grow. 
      Since $\lambda_1$ is a real time-independent eigenvalue the growth of these particular
      solutions is exponential without oscillations. 
      
\item A {\em stable subspace} also of dimension one spawned by the eigenvector associated to the
      eigenvalue with negative real part, ($\lambda_2$).   

			   Solutions in a stable subspace always decay. In our case the eigenvalue associated to this
      subspace, $\lambda_2$, is also a real valued one and therefore the decay of these
      particular solutions is exponential without oscillations.
        
\item A {\em center subspace} which for this range of $k$ is of dimension two and is
      spawned by the eigenvectors associated to the complex eigenvalues with null real part,
      ($\lambda_3$ and $\lambda_4$).
      Solutions in a center subspace oscilate. 

\end{enumerate}

	This case corresponds to the existence of a pair of acoustic modes plus one growing and one
decaying mode. 

	The analysis of the composition of these modes shows that all of them envolve both components
$D$ and $B$, and  are completely coupled through gravity, which, at these large scales plays
an important role. 

\end{enumerate}

	One can then conclude that growth of density perturbations is possible only for $k < k_{_M}$.
	Note that for all $k < k_{_M}$ this growth affects the two components (i.e. the mixture) and
not only one of them.
	This is the reason why we called $k_{_M}$ the Jeans wave number of the mixture. 
	The corresponding physical critical scale length below which no structure could develop is
$l_{_M} = 2\pi/k_{_M}$.

	It is interesting to note that although this scale length is related to the Jeans lengths,
$l_{_B}$ and $l_{_D}$, of the two components when taken separately it does not coincide, in the
general case, with none of them. 
 In fact from relation (\ref{eq.3}) one can see that $l_{_M} = \left( l_{_B}^{-2} + l_{_D}^{-2}
\right)^{-1/2}$.   

	Since we are mainly interested in structure formation we shall consider, in what follows,
only the case when $k < k_{_M}$, (i.e. $l > l_{_M}$).

\section{Behaviour of the general solution} \label{sec.sol}

	In order to study the behaviour with scale of the dynamical system at the onset of the structure
formation process, one needs to analyse the $k$-dependence of the eigenvalues $\lambda_{j}$, the
eigenvectors $\mbox{\boldmath $\xi$}_j$ and the coefficients $\alpha_{j}$, which appear in Eq.
(\ref{eq.1-a}).

 Since $\lambda_2 = -\lambda_1$  and $\lambda_4 = -\lambda_3$ we shall only study the dependence
with scale of the real eigenvalue $\lambda_1$ and of the imaginary part of $\lambda_3$. 

 The eigenvalue $\lambda_1$, being real and positive, is the one related with the growth
rate of the perturbations, is a growing function of scale.
 However $\lambda_1$ does not grow at the same rate for all scales. 
	In fact there are two particular scales at which a significant change in its growth rate
occurs. 
 The first one is the Jeans scale-length of the mixture, $l_{_M}$, where, after changing
from an imaginary to a real value, $\lambda_1$ increases very fast and then slows down,
remaining approximately constant until the second peculiar scale is reached, which occurs
near the Jeans scale-length of the dark matter component, $l_{_D}$, where $\lambda_1$ grows
faster again. 
	Beyond this region the growth of $\lambda_1$ slows down again and approaches the asymptotic
value  $W \! {_{_B}} + W \! {_{_D}}$ as $l \rightarrow \infty$; see Eqs. (\ref{eq.4}) and
(\ref{eq.6}). 

 The eigenvalue $\lambda_3$ is a decaying function of scale and the only change in its
behaviour occurs at a scale which coincides with the second peculiar scale for $\lambda_1$,
(i.e. for a scale near $l_{_D}$), where a faster decrease than for any other region of the
scale range, is observed. 
 For greater scales it approaches zero as $l \rightarrow \infty$.
	Since $\lambda_3$ is the frequency associated to the acoustic modes one can see, from its
dependence with scale, that although oscillations are present in all scales, for large scales
acoustic waves have their frequency approaching zero, i.e. their period tends to infinity. 
	See Fig. \ref{fig.1}.

	Additional information about the system behaviour can be obtained studying the trajectories
in phase-space corresponding to the general solution $\mbox{\boldmath $x$}(t)$;
unfortunately, since the phase-space is of dimension four, they are hard to visualise.
	However, since, as referred above, we are particularly interested in the qualitative
behaviour of the second and fourth components of the general solution $\mbox{\boldmath
$x$}(t)$, i.e. the behaviour of the density perturbations ${\Delta}_{_D}(t)$ and
${\Delta}_{_B}(t)$, we shall infer some properties of the dynamical evolution of the
density perturbations analysing the behaviour of the projection, on the plane $(x_4 , x_2)
\equiv ({\Delta}_{_B} , {\Delta}_{_D})$, of the phase-space four-dimensional trajectories. 

	Looking at the components of the eigenvectors, given by the relations (\ref{eq.4}) -
(\ref{eq.10}), one can see that the projection, on the plane $(x_4 , x_2)$, of the stable
subspace coincides with that of the unstable subspace (since $\beta_{\,1} = \beta_{\,2}$).

	The common projection of these subspaces is the straight line given by the equation:

\begin{equation}
{\Delta}_{_D} = \beta_{\,1} \, {\Delta}_{_B}.
\label{eq.11}
\end{equation}

	The center subspace has a projection on the same plane $({\Delta}_{_B} , {\Delta}_{_D})$
which is also a straight line, defined by

\begin{equation}
{\Delta}_{_D} = \beta_{\,3} \, {\Delta}_{_B}.
\label{eq.12}
\end{equation}

	These equations, (\ref{eq.11}) and (\ref{eq.12}), for the projection of the stable/unstable
and center subspaces can give us the relative composition (relative percentage of components
$B$ and $D$), in the growing and decaying modes, Eq. (\ref{eq.11}), as well as in the acoustic
modes, Eq. (\ref{eq.12}).
	
	From relations (\ref{eq.9}), one can see that since $h^{2} + 4W \! {_{_B}}W \! {_{_D}}$ is
always positive, $\beta_{j}$ are, therefore, real-valued functions for all $k > 0$, with
$\beta_{1} > 0$ and $\beta_{3} < 0$. 
	The condition $\beta_{1} > 0$ means that both components participate positively in the
collapse of the mixture as well as in the decaying mode, which being transient will not
interest us.
 On the other hand $\beta_{3} < 0$ means that, in the acoustic modes, the two components
oscillate with opposite phases; when the density in one of them is growing, the density in the
other is decreasing.

 The direct product of the stable and unstable subspaces forms a two dimensional subspace
which is usually called a {\em saddle subspace} and solutions in such a subspace are
simultaneously pulled towards the equilibrium point along the stable axis and pushed away
from it along the unstable axis. 
 The result is that when the time $t$ varies from $- \infty$ to $+ \infty$ the orbits
representing these particular solutions are initially attracted by the critical point reaching
a closest approach to the origin at a time $t_{cl}$ after which they are repelled away.

 Choosing initial conditions such that $x_1 = x_3 = 0$, i.e. $\mbox{\boldmath $x$}_0$ lies
on the plane $(x_4 , x_2)$, and given the symmetry, in first and third components, between
the eigenvectors $\mbox{\boldmath $\xi$}_{\,1}$ and $\mbox{\boldmath $\xi$}_{\,2}$, one
concludes that the closest approach from the origin of these orbits, is attained at initial
time $t_0$, (i.e. $t_{cl} = t_0$). 
 Therefore, in this case, for $t > t_0$ the system is monotonically driven away from the
origin with time. 
	It is worth noting that this condition ($\mbox{\boldmath $x$}_0$ lying on the $(x_4 ,
x_2)$ plane), corresponds to the reasonable assumption that the mixture starts to collapse
from rest.

 Moreover, since the effect of the center subspace is to force the orbit to oscillate, one
shall have a projected path which oscillates around the projection of the saddle subspace,
i.e. around the straight line given by Eq. (\ref{eq.11}), see Fig. \ref{fig.orb}.
	 
 According to the initial conditions, ${\Delta}_{_B{_{_0}}}$ and ${\Delta}_{_D{_{_0}}}$, one
can see that, if ${\Delta}_{_D{_{_0}}} \! > \beta_{\,3} \, {\Delta}_{_B{_{_0}}}$, the system
will collapse to form a massive structure. 
 On the other hand, if ${\Delta}_{_D{_{_0}}} \! < \beta_{\,3} \, {\Delta}_{_B {_{_0}}}$ it
will evolve to form a void, as represented in Fig. \ref{fig.orb}. 

	Since $\beta_{\,1}$ and $\beta_{\,3}$ depend on the perturbation's wave-number $k$, it is
clear that the direction of the projection of the above mentioned subspaces also changes with
scale, and therefore the relative growth of the perturbations in the two components, as well
as the composition of the collapsing and oscillating modes, is also scale dependent.

 The change in the relative growth of perturbations with scale can be seen analysing the
behaviour of $\beta_{\,1}$ with wave number $k$, or equivalently with scale length $l$.

 The straight lines defined by Eqs. (\ref{eq.11})  and (\ref{eq.12}) make angles with the
direction of the $x_4$-axis, which are given respectively by $\theta_1 = \arctan(\beta_{\,1})$ and
$\theta_3 = \arctan(\beta_{\,3})$. 

	If one assumes that $c_{_B} < c_{_D}$, (case of hot dark matter), the derivative 

\begin{equation}
\frac{d\beta_{\,1}}{dl} = 4\pi^2 \, \frac{c_{_D}^{2} - c_{_B}^{2}}{W \!\! {_{_D}}} \, \left( 1 +
\frac{h}{\sqrt{h^{^2} + 4W \!\! {_{_B}}W \!\! {_{_D}}}} \right) \frac{1}{l^{^3}}
 \label{eq.13-1}
\end{equation}
 is always positive and since $\beta_1(l_{_M}) = c_{_B}^{2}/c_{_D}^{2}$ and  $\lim_{\, l
\rightarrow \infty} \beta_1 = 1$ one concludes that $\theta_1$ is a monotonic function of $l$,
growing from $\arctan(c_{_B}^{2}/c_{_D}^{2})$ until the asymptotic value $\lim_{\, l
\rightarrow \infty} \theta_1 = 45^{\circ}$.

	By a similar study of $\beta_{\,3}$ one concludes that

\begin{equation}
\frac{d\beta_{\,3}}{dl} = 4\pi^2 \, \frac{c_{_D}^{2} - c_{_B}^{2}}{W \!\! {_{_D}}} \, \left( 1 -
\frac{h}{\sqrt{h^{^2} + 4W \!\! {_{_B}}W \!\! {_{_D}}}} \right) \frac{1}{l^{^3}}
 \label{eq.13-a}
\end{equation}
is also always positive and therefore $\theta_3$ grows with scale from $\theta_3(l_{_M}) =
\arctan \left( - \, \frac{ W \!\! {_{_B}} } { W \!\! {_{_D}} } \frac{ c_{_D}^{2} } {
c_{_B}^{2} } \right)$ until the asymptotic value  $\lim_{\, l \rightarrow \infty} \theta_3 =
\arctan \left( - \, \frac{ W \!\! {_{_B}} } { W \!\! {_{_D}} } \right)$.
 This is illustrated in Fig. \ref{fig.subs}.

 It is important to note, however, that the rates of change with scale of both $\beta_{\,1}$
and $\beta_{\,3}$ are not constant. 
	The same is obviously true for $\theta_1$ and $\theta_3$ as shown in the plot of Fig.
\ref{fig.B13}.
	In fact from this plot, or alternatively analysing expressions (\ref{eq.13-1}) and
(\ref{eq.13-a}), one can notice that for $l_{_M} < l \ll l_{_D}$ (i.e. $k_{_M} > k \gg
k_{_D}$), $\beta_{\,1}$ and $\beta_{\,3}$ are slowly increasing functions, while when $l$
reaches the order of $l_{_D}$ they become much steeper functions of $l$. 
	For values of $l$ above this scale the rate of change of $\beta_{\,1}$ and $\beta_{\,3}$
slows down again as $\beta_{\,1}$ and $\beta_{\,3}$ approach the asymptotic values 1 
and $ - \, W \!\! {_{_B}} / W \!\! {_{_D}}$ respectively. 

	This means that the composition of the mixture in the collapsing mode changes substantially
when the perturbation scale is of the order of $l_{_D}$, starting to involve, at this scale,
a much greater part of the dark material which, for smaller scales, was mainly oscillating in
the acoustic mode.

	Further on, it will be shown that there is a critical scale $l_{\rm c}$ where a
resonance between the two components occurs\footnote{This scale $l_{\rm c}$ is in the
particular case we are considering of the order of $l_{_D}$ as shall be seen later. Moreover
it will be shown that, if the mass of hot dark matter particles is $m_{_D} \sim 30\;$eV, the
scale $l_{\rm c}$ has a corresponding mass of the same order as that of typical galaxies.}. 
	It is for this particular scale, and larger ones, that the acoustic modes tend to be devoided
in favour of the collapse mode.

	As a matter of fact, in the limit of very large scales ($k \rightarrow 0$) one has:

\begin{equation}
\lim_{k \rightarrow 0} \frac{ {\Delta}_{_D{_{c}}} } { {\Delta}_{_B{_{c}}} } = 1
 \label{eq.13-b}
\end{equation}
where ${\Delta}_{_D{_{c}}}$ and ${\Delta}_{_B{_{c}}}$ are the density contrast of the dark and
baryonic components participating in the collapse mode. 

 Denoting by  ${\delta}_{_D{_{c}}}$ and ${\delta}_{_B{_{c}}}$ the density perturbations of the
dark and baryonic components in the collapse mode, one has:

\begin{equation}
\lim_{k \rightarrow 0} \frac{ {\delta} \rho_{_D{_{c}}} } { {\delta} \rho_{_B{_{c}}} } =
\frac{\rho_{_D{_{_0}}} } { \rho_{_B{_{_0}}} }
 \label{eq.13-c}
\end{equation}
and for the acoustic mode, by a similar reasoning, one obtains

\begin{equation}
\lim_{k \rightarrow 0} \frac{ {\delta} \rho_{_D{_{a}}} } { {\delta} \rho_{_B{_{a}}} } = -1
 \label{eq.13-d}
\end{equation}
where ${\delta}_{_D{_{a}}}$ and ${\delta}_{_B{_{a}}}$ are the density perturbations in the
dark and baryonic components in the acoustic modes.

	These results mean that, for very large scales, the composition of the perturbed region, in
the collapsing mode, is identical to the one in the unperturbed state, as shown by Eq.
(\ref{eq.13-c}). 
	On the other hand the perturbations in both components, oscillating in the acoustic modes,
cancel each other, since they oscillate with opposite phase and the same amplitude.

\section{The structure formation resonance} \label{sec.res}

	The coefficients $\alpha_{j}$ in Eq. (\ref{eq.1-a}) can be obtained by specifying the
initial conditions $x_{i}(t_{0}) = x_{i_{0}}$. 
 Substituting in the general solution given by Eq. (\ref{eq.1-a}), one obtains:

\begin{equation}
x_{i_{0}} = \sum_{j} \alpha_{j} v_{(j)i}
\label{eq.14}
\end{equation}
where $v_{(j)i}$ is the $i^{th}$ component of eigenvector $j$.

	Solving the above linear system of equations along with the already mentioned assumption
$x_{1_{0}}=x_{3_{0}}=0$ one obtains the coefficients $\alpha_{j}$, which are given by:

\begin{equation}
\begin{array}{l}
 \alpha_{1} =  \frac{ x_{2_{0}} } {2} \; \frac{ 1 - Q_{0} \beta_{\,3} } { \beta_{\,1} -
\beta_{\,3} } \; e^{-\lambda_{1} t_{0}}  \\ 
 \\
 \alpha_{2} =  \frac{ x_{2_{0}} } {2} \; \frac{ 1 - Q_{0} \beta_{\,3} } { \beta_{\,1} -
\beta_{\,3} } \; e^{\lambda_{1} t_{0}} \\
 \\
 \alpha_{3} =  x_{2_{0}} \; \frac{Q_{0} \beta_{\,1} - 1 } { \beta_{\,1} - \beta_{\,3} } \;
\cos{ \left( i \lambda_{3} t_{0} \right) } \\
 \\
 \alpha_{4} =  x_{2_{0}} \; \frac{ 1 - Q_{0} \beta_{\,1} } { \beta_{\,1} - \beta_{\,3} } \;
\sin{ \left( i \lambda_{3} t_{0} \right) } 
 \end{array}
 \label{eq.19}
\end{equation}
where $Q_{0}$ is the ratio between the density contrast in the two components at $t = t_{0}$,
(i.e. $Q_{0} = x_{4_{0}} / x_{2_{0}} = {\Delta}_{_B{_{_0}}} / {\Delta}_{_D{_{_0}}}$), and $i =
\sqrt{-1}$.

	Substituting these expressions in the second and fourth components of the general solution
(\ref{eq.1-a}), one obtains after some straightforward manipulation:

\begin{eqnarray} 
\Delta_{_D}(\tau) & \equiv & x_{2}(\tau) \nonumber \\ 
                  &    =   & x_{2_{0}} \, \left[ \zeta_{1} \left( e^{\lambda_{1}
\tau} +  e^{-\lambda_{1} \tau} \right) + \zeta_{2} \cos (i \lambda_{3} \tau) \right] 
  \label{eq.19-a}
\end{eqnarray}
and
\begin{eqnarray} 
\Delta_{_B}(\tau) & \equiv & x_{4}(\tau) \nonumber \\
                  &    =   & x_{4_{0}} \, \left[ \zeta_{3} \left(
e^{\lambda_{1} \tau} +  e^{-\lambda_{1} \tau} \right) + \zeta_{4} \cos (i \lambda_{3} \tau)
\right] 
 \label{eq.19-b} 
\end{eqnarray}
 where we have used a new time variable, $\tau = t - t_{0}$, by translating the time origin
to the beginning of the process. 
 The new coefficients $\zeta_j$, which are the mode amplitudes for the perturbations in
components $D$ and $B$, are given by the following expressions 

\begin{equation}
\begin{array}{l}
\zeta_{1} = \frac{\beta_{\,1}} {2} \, \frac{ 1 - Q_{0} \beta_{\,3} } { \beta_{\,1} -
\beta_{\,3} } \\
 \\
\zeta_{2} = \beta_{\,3} \, \frac{ Q_{0} \beta_{\,1} - 1 } { \beta_{\,1} - \beta_{\,3}
}\\
 \\
\zeta_{3} = \frac{1} {2} \, \frac{ Q_{0}^{-1} - \beta_{\,3} } { \beta_{\,1} -
\beta_{\,3} } \\
 \\
\zeta_{4} = \frac{ \beta_{\,1} - Q_{0}^{-1} } { \beta_{\,1} - \beta_{\,3}
}
 \end{array}
 \label{eq.19-c}
\end{equation}

 For simplicity it shall be assumed, throughout this section, that $Q_{0}$ is positive,
although massive structures can also originate even in the cases when it is negative, (see the
previous section, Fig. \ref{fig.orb}).

 One must also point out that if one allows $x_{1_{0}} \neq 0$ and/or $x_{3_{0}} \neq 0$ the
expressions (\ref{eq.19}) for $\alpha_{j}$, and consequently the expressions (\ref{eq.19-c})
for $\zeta_{j}$, become more complicated, and therefore dificult to be studied analytically.
 The general analytical study of the mode amplitudes, in the case when all the $x_{i_{0}} \neq
0$, is under development and we intend to publish it in a forthcoming paper.
 However preliminary numerical results seem to indicate that no significant change occurs in
the behaviour of the mode amplitudes.

 Looking at the behaviour of the amplitudes $\zeta_j$, one should first of all point out
that, from Eqs. (\ref{eq.7}), one can see that the pairs $\left( \zeta_{1} \; , \; \zeta_{2}
\right)$ and $\left( \zeta_{3} \; , \; \zeta_{4} \right)$ are related by:    

\begin{equation}
2 \, \zeta_{1} + \zeta_{2} = 2 \, \zeta_{3} + \zeta_{4} = 1
 \label{eq.10a}
\end{equation}

 This means that a maximum of $\zeta_{1}$ corresponds to a minimum of $\zeta_{2}$ and a
maximum of $\zeta_{3}$ corresponds to a minimum of $\zeta_{4}$ or inversely.

 This can be seen in Fig. \ref{fig.3}, where the numerial plot illustrates the dependence of
$\zeta_3$ on scale, and shows its resonant behaviour for $l \sim l_{c}$, and in Fig.
\ref{fig.4},	where the sharp minimum occurring for the coefficient $\zeta_4$ is observed at
the same value of the scale length.
 In the numerical plot of Fig. \ref{fig.coefD} we also illustrate the behaviour of the
coefficients $\zeta_1$ and $\zeta_2$.  
 
	Physically, this means that the occurrence of a resonance (presence of maxima in $\zeta_{1}$
or $\zeta_{3}$), is related with the fact that at the resonant scale the ratio between the
number of particles which populate the collapse and decaying modes and the number of
particles of the same component which populate the acoustic mode grows substantially when
compared with the same ratio in nearby scales.
 Therefore one can say that most of the resonant material is collapsing and little is left
oscillating, with the oscillating mode helping to devoid the background to make a greater
percentage of the particles participate in the collapse. 

 The occurrence of such a resonance can be interpreted as a consequence of the following: 

	As pointed out by Grishchuk \& Zel'dovich (1981), the dynamical system (\ref {eq.1})
represents two gravitationally coupled oscillators.
	If one takes the two components $B$ and $D$ separately, the dispersion relations, for
perturbations in those separate components, give us the natural frequencies of oscillation,
i.e. the ones the components would have if uncoupled, which are given by  

\begin{equation}
\omega_{_B} = \left( c_{_B}^{2} \, k^{2} - W \! {_{_B}} \right)^{1/2}
 \label{eq.19-d}
\end{equation}
and
\begin{equation}
\omega_{_D} = \left( c_{_D}^{2} \, k^{2} - W \! {_{_D}} \right)^{1/2}
 \label{eq.19-e}
\end{equation}
where $\omega_{_B}$ and $\omega_{_D}$ are the natural frequencies for components $B$
and $D$ respectively.

	From Eqs. (\ref{eq.19-d}) and (\ref{eq.19-e}) one can see that the two natural frequencies
coincide for a wave number $k_{\rm c}$ given by

\begin{equation}
k_{\rm c} = \left( \frac{ W \! {_{_D}} - W \! {_{_B}} } { c_{_D}^{2} - c_{_B}^{2} }
\right)^{1/2}.
 \label{eq.19-f}
\end{equation}

	This means that $k_{\rm c}$ is the characteristic wave number for which perturbations in the
two components, taken separately, would have the same collapse time scales.
 Thus it is reasonable to expect a resonance to occur, at least in some of the mode amplitudes
of the perturbations, for a value of $k$ near $k_{\rm c}$. 
	In fact it will be shown below that such a resonance does indeed occur.

	The scale length corresponding to $k_{\rm c}$ is given by

\begin{equation}
l_{\rm c} = 2 \pi \left( \frac{c_{_D}^{2} - c_{_B}^{2}} {W \! {_{_D}} - W \! {_{_B}}}
\right)^{1/2} = \left[ \frac{ l_{_D}^2 - ( W \! {_{_B}} / W \! {_{_D}} ) \, l_{_B}^2 } { 1 -
( W \! {_{_B}} / W \! {_{_D}} ) } \right]^{1/2}. 
 \label{eq.19-g}
\end{equation}

	In the case when $c_{_B} \ll c_{_D}$ and $W \! {_{_B}} \ll W \! {_{_D}}$ one can see that
$l_{\rm c}$ is of the order of $l_{_D}$, as noted in the last section. 

 This resonance has physical meaning only if the value of $k_{\rm c}$ is real and positive.
	Looking at Eq. (\ref{eq.19-f}) one can see that this can only happen in the case when the
gravitationally dominant component, i.e. the one with a greater value of $W$, has also a
greater sound speed, otherwise the scale for which the two components have equal natural
frequencies will be a physical meaningless imaginary one.
	Therefore, in what follows, it will be assumed that $W \! {_{_D}} > W \! {_{_B}}$ and
$c_{_D} > c_{_B}$. 
	This is a necessary condition for the existence of a resonance in the system.
	When applied to our universe, for which the observational evidences seem to indicate to be
dominated by dark matter, (i.e. $W \! {_{_D}} > W \! {_{_B}}$), this means that such a
resonance cannot occur in a cold dark matter (CDM) dominated universe.  
	
	As can be seen in Eqs. (\ref{eq.19-c}), the peculiar behaviour of $\beta_{\,1}$ and
$\beta_{\,3}$ for $l \sim l_{\rm c}$ mentioned in the preceding section, is responsible for
the fact that, from all the perturbations which can start to grow at the recombination epoch,
those with masses of the order of $M_{\rm c}$ are the ones which are going to present, at the
onset of the gravitational instability, the biggest amplitude, in the collapse mode.
	In this sense the scale $l_{\rm c}$ is a privileged one.

 However one must be careful in interpreting this result since it is valid only for the onset
of the instability and cannot be extrapolated in time.
 To have a correct idea about the dynamical evolution of the perturbations one has to solve a
more complicated problem, i.e. we must include, in the equations, the terms due to the
expansion of the universe, which allows us to obtain the correct growth rate of the
perturbations. 
 On the other hand another dificulty arises because we solve the equations only for one
Fourier component of the general solution which separates a perturbation into distinct
wavelengths that, in principle, have distinct growth rates.
 Therefore at any epoch the density contrast in a perturbed region of scale $\sim 1/k$ must
be {\em integrated} in all the region, i.e. for all $k$ in the range $0 < k < 1/k$, (see
Padmanabhan 1993).

 In particular, before a careful analysis of these two points is made, one cannot know which
perturbations are the first to reach the non linear regime. 
 This is a very important question that we shall address in a forthcoming paper (in
preparation).

%  Returning now to the study of the mode amplitudes it can be seen in the plot of Fig.
% \ref{fig.coefD} that $\zeta_{1}$ and $\zeta_{2}$ present also a peculiar behaviour for a
% scale near $l_{c}$.

	Since the growing mode is the one that leads to structure formation we shall be particularly
interested in the behaviour of the coefficients $\zeta_1$ and $\zeta_3$ which are the ones
associated to that mode.

	We shall therefore proceed by analysing the functions $\zeta_{1}$ and $\zeta_{3}$. 
	Looking at their derivatives with respect to $k$, one can see that their maxima occur for
values of $k$ given by:

\begin{equation}
k \, (\zeta_{1}^{\rm max}) = \left[ k_{\rm c}^2 - \frac{2 \, W \! {_{_D}}} {Q_{0} \,
\left( c_{_D}^{2} - c_{_B}^{2} \right)} \right]^{1/2} 
 \label{eq.101}
\end{equation}
for $\zeta_{1}$, and 

\begin{equation}
k \, (\zeta_{3}^{\rm max}) = \left[ k_{\rm c}^2 + \frac{2 \, W \! {_{_B}}} {q \, \left(
c_{_D}^{2} - c_{_B}^{2} \right)} \right]^{1/2} 
 \label{eq.102}
\end{equation}
for $\zeta_{3}$, where $q = 1/Q_{0} = {\Delta}_{_D{_{_0}}} / {\Delta}_{_B{_{_0}}}$.

 Substituting Eq. (\ref{eq.19-f}) into Eq. (\ref{eq.101}) one can see that the existence of
a maximum for $\zeta_{1}$ in the range $0 < k < k_{_{\rm M}}$, can only occur if $q$ satisfies
the following condition:

\begin{equation}
q < \frac{1}{2} \, \left( 1 - \frac {W \! {_{_B}}} {W \! {_{_D}}}
\right)
 \label{eq.102a}
\end{equation}

  On the other hand, substituting Eq. (\ref{eq.19-f}) into Eq. (\ref{eq.102}) one can see that
the existence of a maximum for $\zeta_{3}$ in the same range, can only occur if:

\begin{equation}
q > 2 \, \left( \frac {c_{_D}^{2}} {c_{_B}^{2}} - \frac {W \! {_{_D}}} {W \!
{_{_B}}} \, \frac {c_{_B}^{2}} {c_{_D}^{2}} \right)^{-1}
 \label{eq.102b}
\end{equation}

	It is reasonable to assume that the composition of the initial perturbation is the same as
the composition of the unperturbed state (i. e.  $q = W \! {_{_D}}/W \! {_{_B}}$) .
 In this case, it is easy to see that $\zeta_{1}$ will never have a maximum in the desired
range, while $\zeta_{3}$ can have one, only if the following condition is satisfied:

\begin{equation}
 \frac {c_{_D}^{2}} {c_{_B}^{2}} > \frac {W \! {_{_B}}} {W \! {_{_D}}} + \sqrt{ \frac {W \!
{_{_B}}^2} {W \! {_{_D}}^2} + \frac {W \! {_{_D}}} {W \! {_{_B}}} }. 
 \label{eq.102c}
\end{equation}

 From this analysis one concludes that, in a hot dark matter dominated-universe, the
resonance will only affect the baryonic component. 
 We shall, therefore, from now on study only the behaviour of the mode amplitude $\zeta_{3}$
which is the one associated to the growing mode in the baryonic component.

Substituting the value of $k \, (\zeta_{3}^{\rm max})$ given by Eq. (\ref{eq.102}) into  Eq.
(\ref{eq.19-c}), one obtains for the function $\zeta_{3}$ its maximum value $\zeta_{3}^{\rm
max}$, given by: 

\begin{equation}
\zeta_3^{\rm max} = \frac{1}{4} \left( 1 + \sqrt{1 + \frac{W \! {_{_D}}} {W \!
{_{_B}}} \, q^2 } \; \right) 
 \label{eq.103}
\end{equation}
which depends only on the unperturbed densities of the components and on the initial
conditions.

	In the case of a universe dominated by dark matter ($W \! {_{_D}}/W \! {_{_B}} \gg 1$) and
for the initial conditions specified above, one can say that $\zeta_{3}^{\rm max} \simeq
\frac{1}{4} \, \left( \frac{W \! {_{_D}}}{W \! {_{_B}}} \right)^{3/2}$. 

 From this expression it is clear that the height of the resonant peak increases as the
parameter $q$, (in this case of the order of $W \! {_{_D}}/W \! {_{_B}}$), increases.
	This means that the more the Universe is dominated by dark matter, the more the effect of the
structure formation resonance in the baryonic material will be important.

	In Fig. \ref{fig.3} the amplitude $\zeta_3$, of the growing mode in the baryonic component,
is ploted as a function of the mass scale of the perturbation, for $q = 19$, which corresponds
to assume that $q = W \! {_{_D}}/W \! {_{_B}} = \Omega_D/\Omega_B$, using for $\Omega_B$ and
$\Omega_D$ the values $0.05$ and $0.95$ respectively.
 This is to be compared with the plots in Fig. \ref{fig.multzeta3} where other values of the
parameter $q$ were used.  

	The importance of the resonance can be also measured by the width of its band: a larger
resonant band corresponds to a weaker resonance (Feynman 1966). 
 We shall define the resonant band width $\gamma$ as the width of the curve of
$\zeta_3^{\,^2}$ at half the maximum of its height.
 
 The two values of $k$ ($k_{1,2}$), which define the resonant band width $\gamma = \left|
k_{1} - k_{2} \right|$, must therefore satisfy the following condition: 

\begin{equation}
\zeta_3^{\,^2} (k_{1,2}) = \frac{1}{2} \left( \zeta_3^{\rm max} \right)^2 
 \label{eq.103a}
\end{equation}
and are given by

\begin{equation}
k_{1,2} = \left[ k_{\rm c}^2 + \frac{2 \, W \! {_{_D}} \, q} {d} \pm \left(
\sqrt{2} - 4 \, \zeta_3^{\rm max} \right) \frac{ \sqrt{r} } {d} \; \right]^{1/2} 
 \label{eq.104}
\end{equation}
where

\begin{equation}
d = 4 \, \zeta_3^{\rm max} \left( 2 \, \zeta_3^{\rm max} - \sqrt{2} \right) \,
\left( c_{_D}^{2} - c_{_B}^{2} \right)   
 \label{eq.105}
\end{equation}
and

\begin{eqnarray}
r & = & \left[ {W \! {_{_D}}}^{\!\! 2} \, q^2 + 2 \left( \sqrt{2} - 1 \right) W \! {_{_B}} W
\! {_{_D}} + \right. \nonumber \\
  &   & \left. 2 \left( \sqrt{2} - 1
\right) \sqrt{ W \! {_{_B}} W \! {_{_D}} \left( {W \! {_{_D}}}^{\!\! 2} \, q^2 + W \! {_{_B}}
W \! {_{_D}} \right) } \; \right]^{1/2}   
 \label{eq.106}
\end{eqnarray}

	For given sound speeds of the two components one can see that this width is very sensitive to
the product $W \! {_{_B}} W \! {_{_D}}$.
	On the other hand, the further apart the sound speeds $c_{_D}$ and $c_{_B}$ in the two
components are, the smaller the resonant band width will be, indicating a stronger and sharper
resonance.

	Another feature one should point out is the behaviour of $\zeta_3$ to the left and to the
right of the resonant scale.

	Since there is no growing mode for $k > k_{_{\rm M}}$, we shall be interested in studying the
behaviour of $\zeta_3$ for the limits $k \rightarrow 0$ and  $k \rightarrow
k_{_M}$. 

 The value of $\zeta_3$ in the limit $k \rightarrow 0$, depends on the values of $W \!
{_{_B}}$ and $W \! {_{_D}}$ and is given by:
 
\begin{equation}
\lim_{k \rightarrow 0}{\zeta_3} = \frac{1}{2} \; \frac{q \, W \! {_{_D}} + W \!
{_{_B}}}{W \! {_{_D}} + W \! {_{_B}}} 
 \label{eq.109}
\end{equation}
whereas the value of $\zeta_3$ for $k = k_{_{\rm M}}$ depends on both the sound speeds and on
the densities of the two components, and is given by:

\begin{equation}
\zeta_3 (k_{_M}) = \frac{1}{2} \; \frac{q + \frac{W \! {_{_B}}}{W \! {_{_D}}} \,
\frac{c_{_D}^{2}}{c_{_B}^{2}}} {\frac{c_{_B}^{2}}{c_{_D}^{2}} + \frac{W \! {_{_B}}}{W
\! {_{_D}}} \, \frac{c_{_D}^{2}}{c_{_B}^{2}}}
 \label{eq.200}
\end{equation}

 Assuming, as above, that $q \sim W \! {_{_D}}/W \! {_{_B}}$, the asymptotic value $\lim_{k
\rightarrow 0}{\zeta_3} \gg 1/2$, as can be seen from Eq. (\ref{eq.109}), while, if one allows
$q$ to be much smaller than $1$, then $\lim_{k \rightarrow 0}{\zeta_3} \ll 1/2$.

 On the other hand, in the case $W \! {_{_D}}/W \! {_{_B}} \ll c_{_D}/c_{_B}$, and for
reasonable values of $q$, the value of $\zeta_3 (k_{_M})$ is always of the order of $1/2$.

	One can therefore conclude that the $\zeta_3$ curve has a different behaviour on the far
left and on the far right of the resonant scale.  
 For small scales the value of $\zeta_3$ does not vary very much with initial conditions,
while for great scales that value is very sensitive to initial conditions.
	The smaller the initial perturbation in the dark component when compared with the initial
perturbation in the baryonic component, the smaller will be the value of the mode amplitude in
the growing mode of the baryonic component. 
 For scales larger than the resonant one, favouring, in consequence, the onset of
perturbations at the resonant scale or smaller (relative to greater ones).
 The inverse occurs if the perturbation in the dark component increases relative to the
perturbation in the baryonic component, i.e. in this case the favoured scales for the onset of
perturbations are the resonant one or greater (see Fig.\ref{fig.multzeta3}).

\section{The resonance, the galactic scale and the mass of dark matter particles}
\label{sec.gxf}

 We believe that the resonance described in Sect. \ref{sec.res} is responsible for setting
up the growth, at the recombination epoch, of massive structures at the galactic scale.
 The resonance therefore provides the mechanism to select out this particular scale, among
all possible scales which could undergo the gravitational collapse.
 In other words this resonance provides an explanation why do galaxies, with the typical
masses one observes, form around the recombination epoch, even if the Jeans mass for
baryons, and even the Jeans mass for the mixture, either before or just after this epoch, is
either much larger or smaller than typical galactic masses. 

	The typical galactic masses should, according to this work, correspond to the mass of the
perturbations, which, at the recombination epoch, would have a scale near the above mentioned
resonant scale.
	One would therefore have no need to assume that the mass spectrum of perturbations at that
time was already populated in a way similar to the one we observe now.
 Instead, one can say that the typical galactic mass scale would simply result from the
scale selected by this resonance.
 
 Of course we need a better and deeper understanding of the physics of the resonant mechanism
and how it can affect the power spectrum of the density perturbations, not only at
recombination epoch but also in the period following it.
 Therefore one has to go from separate Fourier amplitudes to quantities {\em integrated} over
the wavelength scale. 
 Since one is interested in knowing how the spectrum of inhomogeneities will transform after
recombination, one must follow the dynamics of several Fourier components of the density
perturbations from recombination until the epoch when the perturbations become non linear.

 Note that in the matter dominated period, before recombination, no resonance can exist.
 This is because, due to the coupling between baryonic matter and radiation in that period,
the component with greater sound speed is the baryonic one. 
 On the other hand, in a flat universe like the one we have assumed, the baryons are not
gravitationally dominant. 
 Therefore one can see from Eq. (\ref{eq.19-f}) that the resonance cannot occur in that
period.

 Thus, in order to have a detailed understanding of galaxy formation one only needs to study
the dynamics of the Fourier components after recombination epoch.
 Before recombination one can assume that the power spectrum was any reasonable one predicted
by another model, like, for instance, some type of inflationary model.     

 Since this is a very important point which must be carefully analysed. 
 Since it lies out of the main goal of this paper, we intend to treat it thoroughly in a
separate paper. 

	One can relate the scale of the resonance with the mass of the dark matter particle. 
 The scale at which the resonance peak occurs depends on $c_{_D}$, see Eq. (\ref{eq.102}),
which in turn is a function of the value of the mass of dark matter particles $m_{_D}$. 
 From Eq. (\ref{eq.19-f}) we have already concluded that in order for the resonance to occur,
the dark matter must necessarily be hot.

	The total mass $M$ inside a spherical region of radius $l/2 = \pi/k$ is given by 

\begin{equation}
M = \frac{4}{3} \, \pi^4 \, \frac {\rho} {k^3}, 
 \label{eq.201}
\end{equation}
where $k$ is the wave number and $\rho = \rho_{_B} + \rho_{_D}$ is the total mean density in
that region.

 In order to obtain bounds for the mass of dark matter particles we shall assume that galaxies
form by means of the above described resonance and therefore have mass scales which correspond
to the resonant scale for $\zeta_{3}$.
	The range of galactic masses which one observes in the universe will therefore imply a
specific range for the mass of hot dark matter particles.

 Substituting in Eq. (\ref{eq.201}) the value of the wave number $k$ by its resonant value $k
\, (\zeta_{3}^{\rm max})$, given by Eq. (\ref{eq.102}), one can establish the following
relation between $M$ (in solar masses) and $m_{_D}$ (in eV): 

\begin{equation}
m_{_D} \simeq 210 \, \pi^{5/6} \; \frac{\rho^{1/3}}{\rho_{_D}^{1/2} M^{1/3}}
 \label{eq.202}
\end{equation}

 In deriving this approximate equation\footnote{Although an exact, but more complicated,
expression can be derived for the mass of dark matter particles, the use of these
assumptions, which are valid for the case of HDM, allows us to derive a simple but accurate
expression for $m_{_D}$.}, we have assumed that $W \! {_{_D}} \gg W \! {_{_B}}$ and that
$c_{_D} \gg c_{_B}$. 
 The sound speeds $c_{_j}$ at recombination, appearing in Eq. (\ref{eq.102}) are given by Eq.
(\ref{eq.13}). 

	One should emphasize that from the decoupling, between the neutrino-like particles and
radiation, until the epoch at which these particles become non-relativistic, which for $m_{_D}
\sim 30 \,$eV occurs at a red-shift $z_{_{NR}} \sim 6 \times 10^4$, (Doroshkevich et al.
1988), their temperature T$_{D}$ is given by T$_{D} = (4/11)^{1/3}\,$T$_{\gamma}$, where
T$_{\gamma}$ is the radiation temperature. 
 Since then, it decays with $a^{-2}$, where $a$ is the scale factor of the universe (Gao \&
Ruffini 1981; Padmanabhan 1993).
	One can therefore estimate that, at recombination, such a dark component $D$ will have a
temperature T$_{D{_{rec}}} \sim 40 \, $K. 

 In Fig. \ref{fig.mD} is plotted the scale $M$ at which the resonant peak occurs (which we
have assumed to be the galactic total mass) versus the mass of hot dark matter particles
$m_{_D}$. 

	Assuming that a typical galaxy formed at recombination epoch has a baryonic mass of a few
$10^{11}\,$M$_{\odot}$ one can estimate from the plot in Fig. \ref{fig.mD} that the mass of
the dark matter particles is of the order of $30\,{\rm eV}$. 
 It is remarkable that this value agrees quite well with the cosmological upper bounds for
neutrino's masses (Peebles 1993) as well as the values predicted by Sciama's Decaying Dark
Matter Hypothesis (Sciama 1990a; Sciama 1990b).

\section{Conclusions} \label{sec.con}

	It is clear from the results obtained in this work that it is incorrect to use the Jeans
criterion for a 1-component fluid to a 2-component mixture, for two main reasons:

\begin{enumerate}

\item The Jeans scale for the mixture does not correspond to any of the Jeans scales for
      the components taken separately.

\item Moreover, in a 2-component mixture there are always present, at all scales,
      oscillations in the two components, which correspond to acoustic waves. The same is
      not true for a one component system.

\end{enumerate}

	However the main result of this work is, as we have shown, that, in a two-component fluid, a
resonance between the two components of the fluid must always occur provided that the
component with a greater density is the one with a greater sound speed, see Eq.
(\ref{eq.19-f}). 
	One can therefore put some constraints on the nature of the dark matter if galaxies are to be
formed by such a resonant effect.
	
 In fact, assuming that the density parameter of the universe is $\Omega = 1$ and adopting a
value for the density of the baryonic component in agreement with standard light-element
nucleosynthesis, one is led to the conclusion that the universe is dominated by
dark matter, i.e. $W \! {_{_D}} > W \! {_{_B}}$.
	In such a case, if $c_{_B} > c_{_D}$ (case of CDM), one concludes from Eq. (\ref{eq.19-f})
that no resonance between the two components is possible and therefore, the galactic scale
would not be a preferred one, as it seems to be. 
 One is therefore led to believe that most, if not all, of the {\em dark matter present in
the universe is hot}, i.e. it is constituted by light neutrino-like particles as
implied by the condition for the existence of a strong resonance, i.e. $c_{_D} \gg c_{_B}$,
(see Eq. (\ref{eq.104}) and comments following it).

	This resonance is linked to the fact that at the critical scale $l_{c}$, there is an
enhancement of the number of particles in both components, (but particularly in the baryonic
one), which participate in the collapse mode, instead of participating in the acoustic modes
as it happens for other scales.
	This effect is only significant in a narrow resonant band in the length, or mass, scale of
the perturbations.
	This, we believe, is the reason for the formation of galactic structures at the time of
recombination with their characteristic range of mass scales.  

 This effect occurs at the galactic scales provided that the dark matter component is made of
neutrino-like particles with masses of the order of $30\,$eV.
	This value for the mass of the neutrino-like particles is below the upper bounds imposed by
cosmological constraints on the neutrino's mass (Peebles 1993), assuming a flat universe,
and remarkably close to the value predicted by Sciama's Decaying Dark Matter Hypothesis
(Sciama 1990a; Sciama 1990b).
	We must also note that, from the numerical results obtained in the preceding sections, this
mass scale is of the order of $10^{13}$M$_{\odot}$ which gives a baryonic mass, in the same
scalelength, of the order of $\frac{ \; \rho_{_B} } { \rho_{_D} } 10^{13}$M$_{\odot}$, which
lies in the range of the masses of typical galaxies (a few $10^{11}$M$_{\odot}$).

	One should point out, however, that the resonant scale $l_{c}$ is particularly sensitive to
the mass of the particles constituting the dark component, see Fig. \ref{fig.mD}.

\section{Future work} \label{sec.fut}

 This study has left a lot of open questions, some of which are already under investigation.
 Below we briefly comment on some of them.
 
 First: One should point out that this study assumes galaxies to have started their formation
during recombination.
 However, this can well be not the case and they could have started their formation over a
wider time interval starting from recombination.
 In that case, the unperturbed densities $\rho_j$ and the sound speeds $c_j$ in the two
components would vary during such a period, allowing the scale of the resonance to be shifted.
 This might in turn provide an explanation for the observed wide mass range for galactic
structures. 
 This is a problem which we intend to address in a future paper (in preparation), where the
effects due to the expansion of the universe are taken into account. 

 Second: Although the analysis of possible effects of this resonant mechanism on the power
spectrum is out of the main purpose of this paper, the existence of the above mentioned
resonance, or at least the change in behaviour of the mode amplitudes near the resonant
scale is, on its own, a very important result.
 The occurrence of this resonance can have, in principle, important implications in the
process of galaxy formation, and, in particular, transform the power spectrum of density
perturbations at recombination epochs.
	Therefore, though a better understanding of the physics underlying the resonant mechanism is
needed, this point shall be thoroughly treated in a forthcoming paper of ours.

	Third: One must note that it will be interesting to study what happens for the case of a
3-component system like the dark matter mixed models.
	We suspect that in such a case more than one structure formation resonance occurs.
	This is a question which we intend to address in the future. 

 Finally, one should also mention that although resonances between periodic coupled
oscillators are common in all branches of physics, this type of resonance occurring in (non
periodic) collapsing modes are, to our knowledge, yet unheard of, and we suspect that in other
branches of physics such as in plasmas (where one has a 2-component fluid whose components
are coupled by the electromagnetic interaction) similar resonances may perhaps occur, giving
rise to instabilities. 
 We think that such an open question is also worth investigating.

\begin{acknowledgements}

 We would like to thank the referee (Dr. M. Tagger) for his helpful criticism which strongly
helped us to clarify some important results presented in this work.

	We also thank JNICT for financial support through its STRIDE projects program, and Centro de
Astrof\'{\i}sica da Universidade do Porto (CAUP) for the facilities  provided.

\end{acknowledgements}

\newpage

\begin{figure}[htb]
%\hspace{2.0cm}
%\special{illustration L13} 
 \caption[]{Dependence on scale (in Solar Masses), of the eigenvalue $\lambda_1$ and imaginary
part of $\lambda_3$. Note the two relevant mass scales for $\lambda_1$ and only one for
$\lambda_3$ which coincides with the second one for $\lambda_1$. The vertical arrow indicates
the total mass $M_{_D}$ inside a spherical region of diameter $l_{_D}$}     
 \label{fig.1} 
\end{figure}

\begin{figure}[htb]
%\hspace{2.0cm}
%\special{illustration L13} 
 \caption[]{Typical projected phase-space trajectories, representing the general solution of the
dynamical system, on the plane $(x_4 , x_2)$. Also plotted are the projections of center and
saddle subspaces. The left hand orbit corresponds to the situation ${\Delta}_{_D{_{_0}}} \! <
\beta_{\,3} \, {\Delta}_{_B{_{_0}}}$ and the right hand one to the inverse initial condition.
(This particular plot corresponds to a mass scale of the perturbation of $\sim
10^{13}$M$_{\odot}$)} 
 \label{fig.orb}   
\end{figure}

\begin{figure}[htb]
%\hspace{2.0cm}
%\special{illustration L13} 
 \caption[]{Projection on the $(x_4 , x_2)$ plane of the three subspaces described in text for
$l > l_{_M}$. The long-dashed lines represent the projection of the saddle subspace, in the
limits $l = l_{_M}$ and $l \rightarrow \infty$. The short-dashed ones are the projections of
the center subspace for the same limits}    
 \label{fig.subs} 
\end{figure}

\begin{figure}[htb]
%\hspace{2.0cm}
%\special{illustration L13} 
 \caption[]{Scale dependence of the angle between the projection of the saddle and center
subspaces with the $x_4$-axis, respectively $\theta_1$ and $\theta_3$. The vertical arrow
indicates the total mass $M_{_D}$ inside a spherical region of diameter $l_{_D}$. Note the
very peculiar behaviour around $M_{_D}$}    
 \label{fig.B13}
\end{figure}

\begin{figure}[htb]
%\hspace{2.0cm}
%\special{illustration L13} 
 \caption[]{Coefficient $\zeta_3$ and its dependence on scale. The vertical arrow indicates the
total mass $M_c$ inside a spherical region of diameter $l_{c}$. Note the clear change in
behaviour of this coefficient around this scale}    
 \label{fig.3}
\end{figure}

\begin{figure}[htb]
%\hspace{2.0cm}
%\special{illustration L13} 
 \caption[]{Coefficient $\zeta_4$ and its dependence on scale. See Eq. (39) and the
discussion following it. The vertical arrow indicates the total mass $M_c$ inside a spherical
region of diameter $l_{c}$}    
 \label{fig.4}
\end{figure}

\begin{figure}[htb]
%\hspace{2.0cm}
%\special{illustration L13} 
 \caption[]{Coefficients $\zeta_1$ and $\zeta_2$ and their dependence on scale. Note that the
behaviour of these mode amplitudes, although not similiar to the one of $\zeta_3$ and
$\zeta_4$, is also peculiar for a scale around $l_{c}$. The vertical arrow indicates the
total mass $M_c$ inside a spherical region of diameter $l_{c}$}    
 \label{fig.coefD}
\end{figure}

\begin{figure}[htb]
%\hspace{2.0cm}
%\special{illustration L13} 
 \caption[]{Dependence on scale (in Solar Masses), of the growing mode amplitude $\zeta_3$, for
distinct initial conditions described by different values of the parameter $q$. One can
notice that for $q > 1$ the values of $\zeta_3$ to the right of the resonant scale are higher
than their values to the left of it. The opposite occurs for $q < 1$}     
 \label{fig.multzeta3} 
\end{figure}

\begin{figure}[htb]
%\hspace{2.0cm}
%\special{illustration L13} 
 \caption[]{Dependence of the scale $M$, (in Solar Masses), at which the resonant peak in the
amplitude $\zeta_{3}$ occurs, plotted as a function of the mass $m_{_D}$ of the particles 
which constitute hot dark matter}     
 \label{fig.mD} 
\end{figure}

%\newpage

%\listoffigures


\begin{thebibliography}{}

 \bibitem{Arnold73}
Arnold V. I., 1973, Ordinary Differential Equations, MIT Press, Cambridge MA
 \bibitem{Doros88}
Doroshkevich A.G., Klypin A.A., Khlopov M.Yu 1988, SvA, 32(2), 127
 \bibitem{Feynman 66}
Feynman R.P., Leighton R.B., Sands M., 1966, The Feynman Lectures on Physics, Vol.I,
 Addison-Wesley Pub. Comp., Massachusetts 
 \bibitem{GaoRuf81}
Gao J. G., Ruffini R. 1981, Phys. Lett., 100B, 47 
 \bibitem{Grishchuk81}
Grishchuk L. P., Zel'dovich Ya. B. 1981, Sov. Astron., 25(3) 
 \bibitem{Jeans02}
Jeans J. 1902, Phil. Trans. R. Soc., 199A, 49 
 \bibitem{Jeans28}
Jeans J. 1928, Astronomy and Cosmogony, Cambridge University Press, Cambridge 
 \bibitem{Kolb}
Kolb E., Turner M. 1990, Early Universe, Addison-Wesley, Massachusetts 
 \bibitem{Pada}
Padmanabhan T. 1993, Structure Formation in the Universe, Cambridge University Press,
 Cambridge  
 \bibitem{Peebles 93}
Peebles P. J. E., 1993, Principles of Physical Cosmology, Princeton Series in
 Physics, Princeton University Press, Princeton
 \bibitem{Polyachenko81}
Polyachenko V. L., Fridman A. M. 1981, Sov. Phys. JETP, 54(1)
 \bibitem{sciama90a}
Sciama D. W. 1990a, MNRAS 246, 191
 \bibitem{sciama90b}
Sciama D. W. 1990b, Nat 348, 617
 \bibitem{Walker et al. 91}
Walker T. P., Steigman G., Schramm D. N., Olive K. A., Kang H., 1991, Ap.J., 376, 51 
 \bibitem{Weinberg72}
Weinberg S. 1972, Gravitation and Cosmology, J. Wiley, New York
 \bibitem{Zel'dovich83}
Zel'dovich Ya. B., Novikov I. D. 1983, Relativistic Astrophysics (Vol. 2), The University of
 Chicago Press, Chicago 

\end{thebibliography}
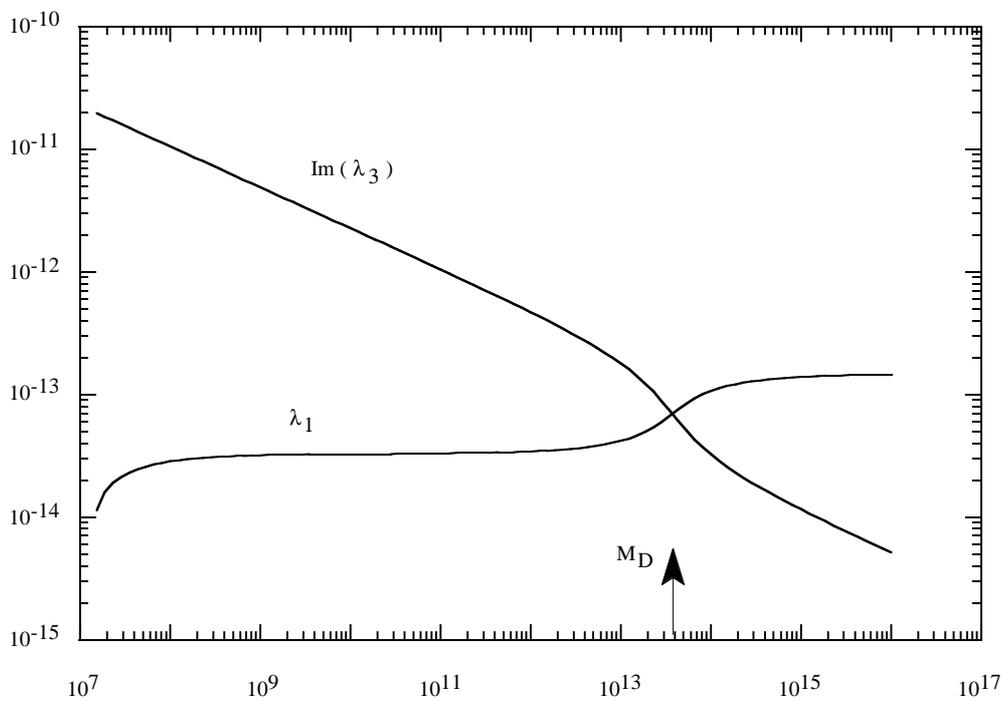
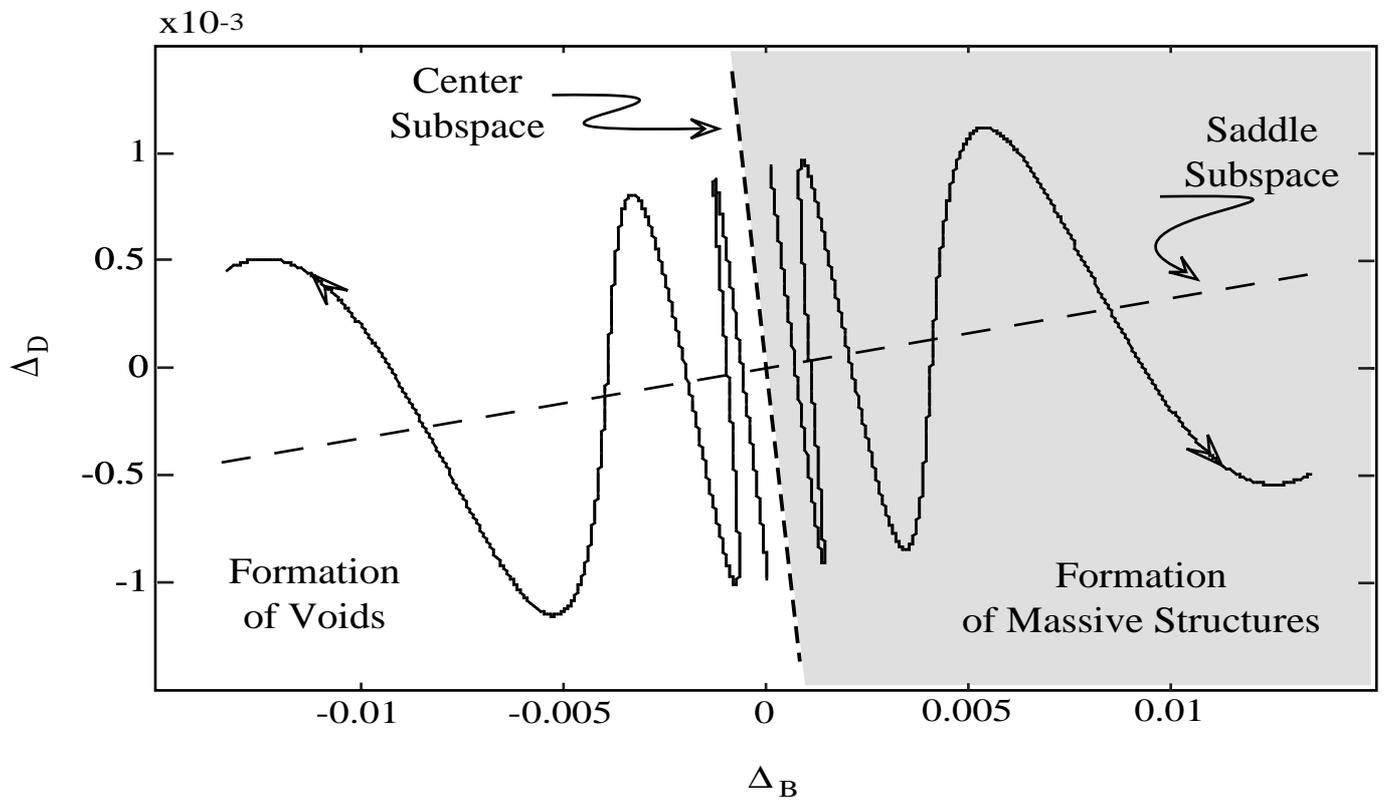
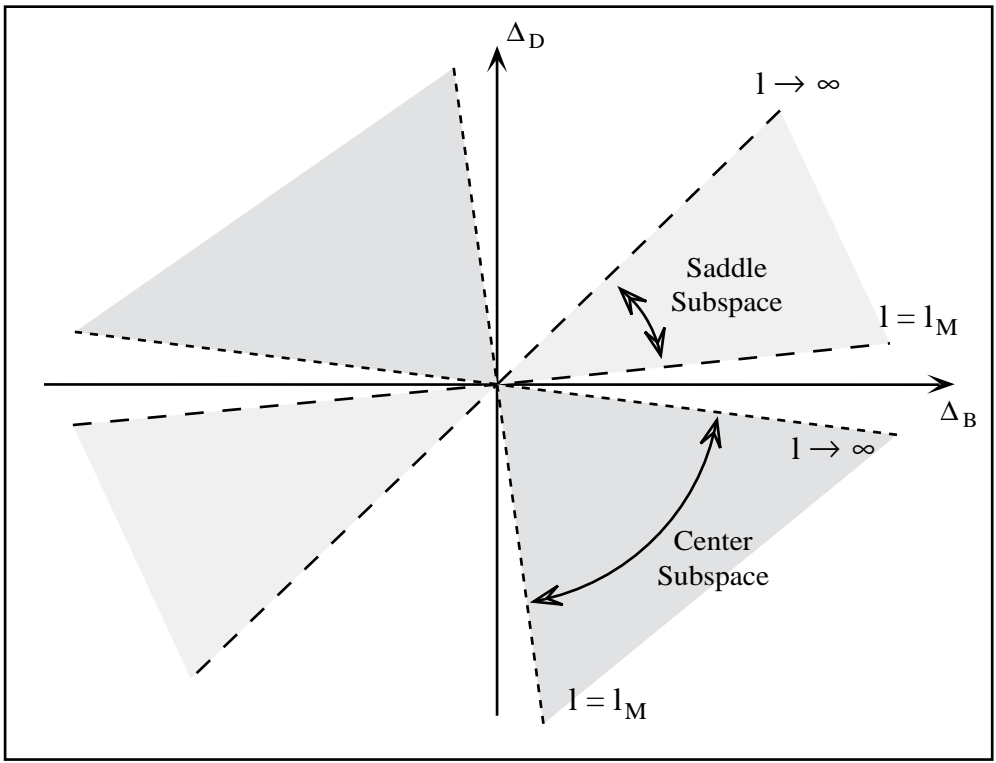
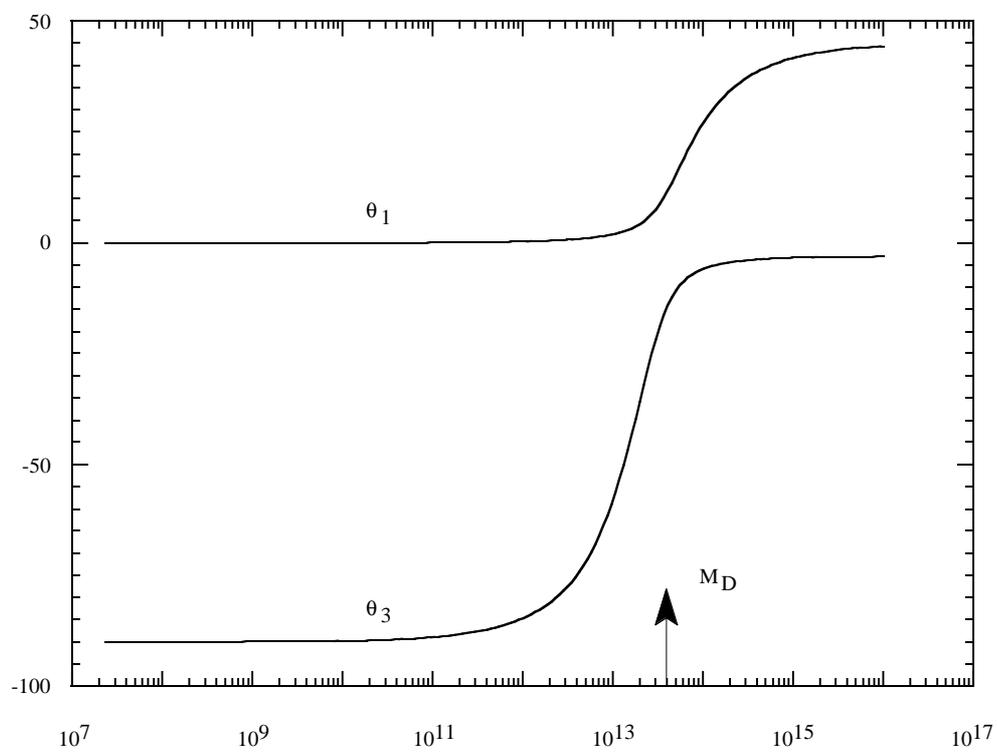
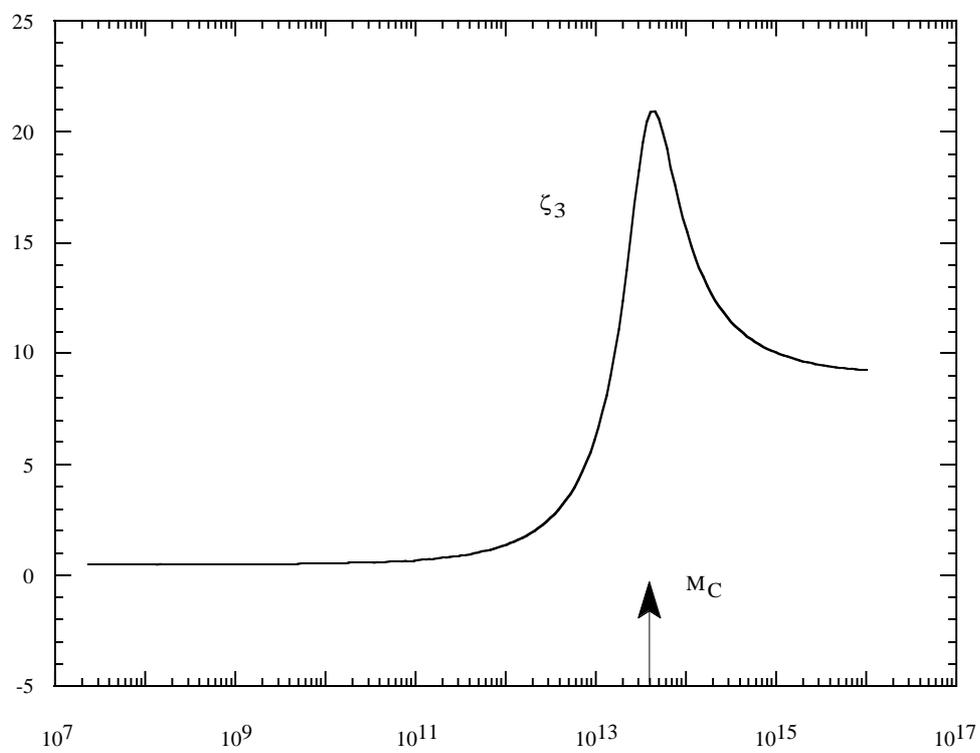
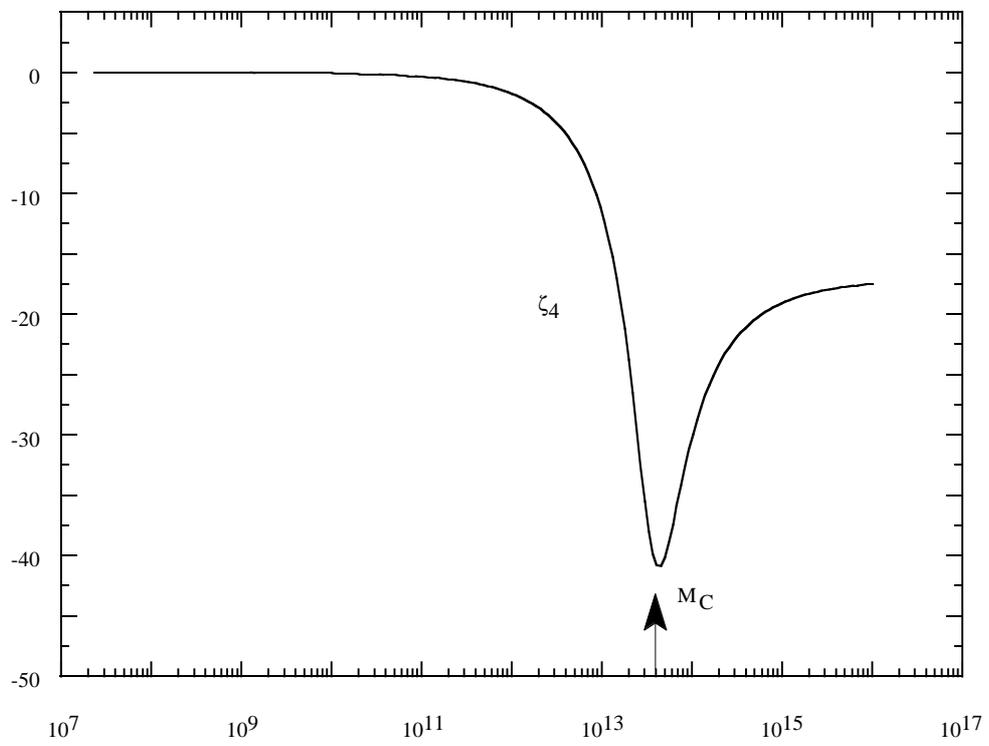
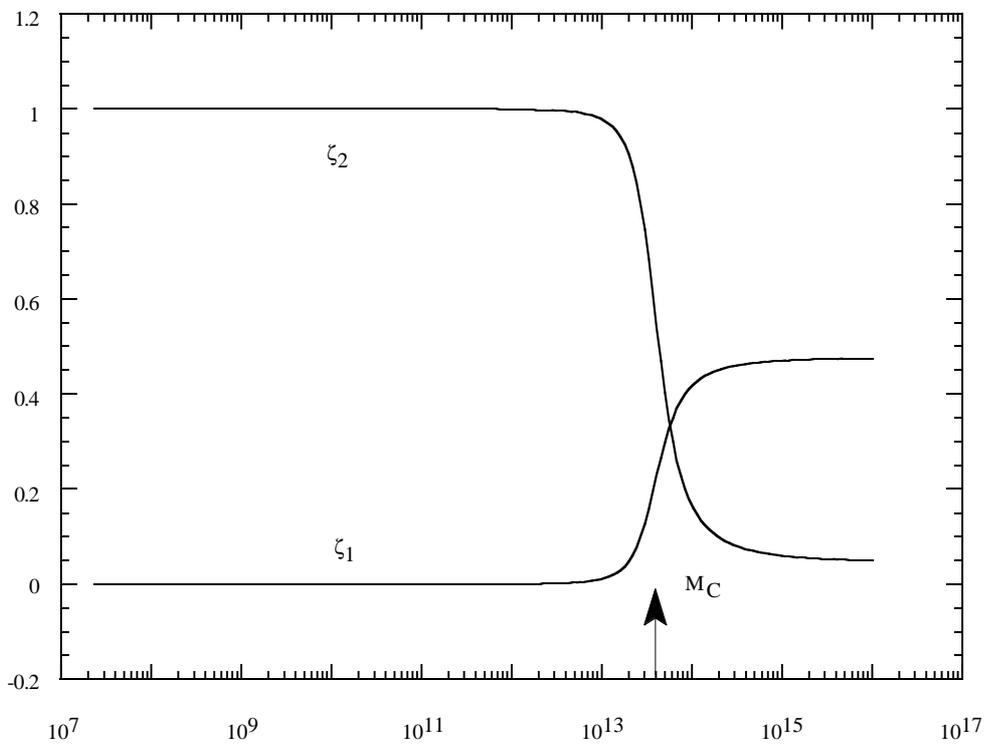
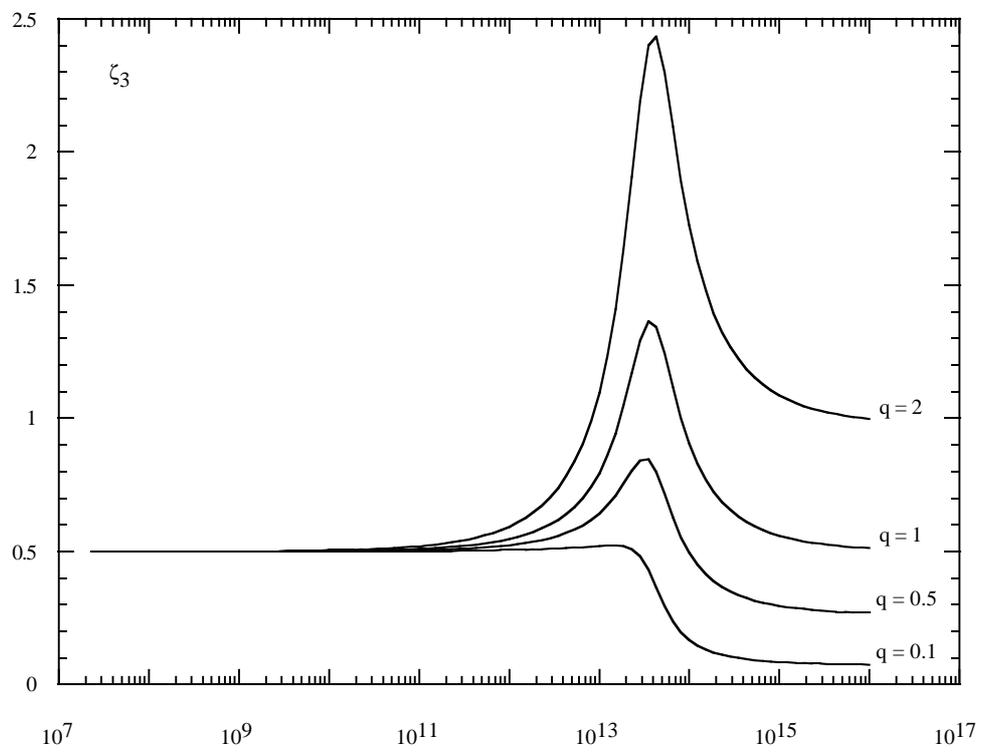
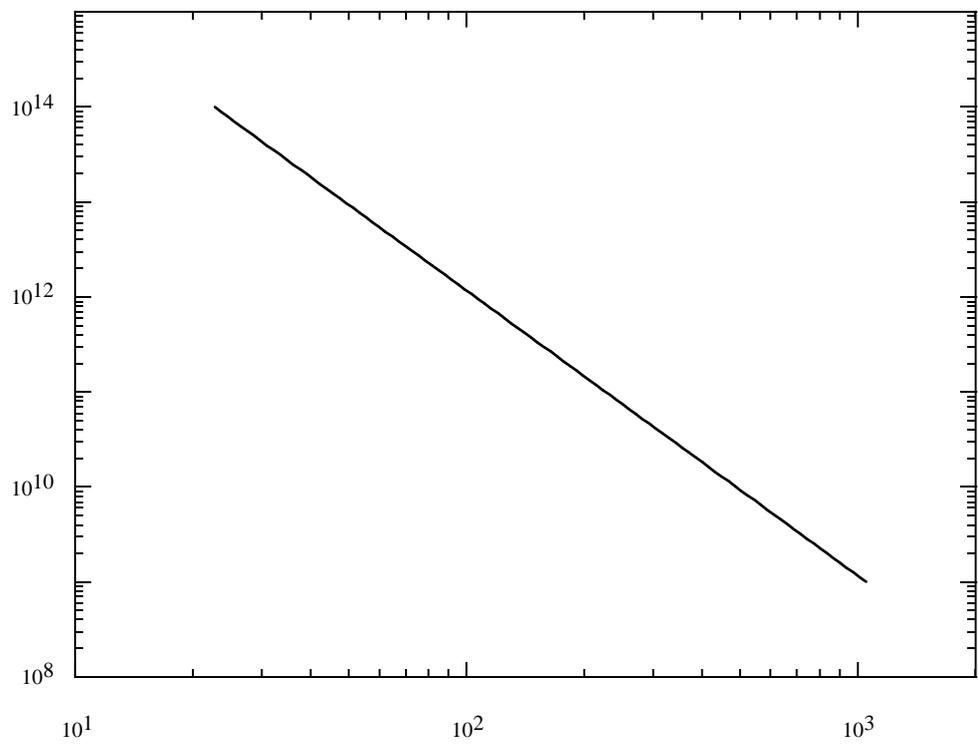
\end{document}